# Towards Effective Collaboration between Software Engineers and Data Scientists developing Machine Learning-Enabled Systems

Gabriel Busquim
gbusquim@inf.puc-rio.br
Pontifical Catholic University of Rio de Janeiro
Rio de Janeiro, Brazil

Allysson Allex Araújo
allysson.araujo@ufca.edu.br
Federal University of Cariri
Ceará, Brazil

Maria Julia Lima
mjulia@tecgraf.puc-rio.br
Pontifical Catholic University of Rio de Janeiro
Rio de Janeiro, Brazil

Marcos Kalinowski
kalinowski@inf.puc-rio.br
Pontifical Catholic University of Rio de Janeiro
Rio de Janeiro, Brazil

## ABSTRACT

Incorporating Machine Learning (ML) into existing systems is a demand that has grown among several organizations. However, the development of ML-enabled systems encompasses several social and technical challenges, which must be addressed by actors with different fields of expertise working together. This paper has the objective of understanding how to enhance the collaboration between two key actors in building these systems: software engineers and data scientists. We conducted two focus group sessions with experienced data scientists and software engineers working on real-world ML-enabled systems to assess the relevance of different recommendations for specific technical tasks. Our research has found that collaboration between these actors is important for effectively developing ML-enabled systems, especially when defining data access and ML model deployment. Participants provided concrete examples of how recommendations depicted in the literature can benefit collaboration during different tasks. For example, defining clear responsibilities for each team member and creating concise documentation can improve communication and overall performance. Our study contributes to a better understanding of how to foster effective collaboration between software engineers and data scientists creating ML-enabled systems.

## 1 INTRODUCTION

The processes behind developing and maintaining Machine Learning (ML) solutions have raised considerable challenges for the Software Engineering (SE) processes [17]. In this study, we refer to systems with ML components as ML-enabled systems, as their behavior is dictated both by explicitly defined rules and the data used by the ML model [16]. One particular challenge relies on the fact that teams working on ML-enabled systems are often multidisciplinary, where actors with different roles must collaborate in an orchestrated manner during their technical tasks [1]. More specifically, the development and integration of ML models with other system components usually involves collaboration between two key actors: Software Engineers (SWEs) and Data Scientists (DSCs).

As clarified by Lewis *et al.* [11], an ineffective collaboration between SWEs and DSCs can result in misunderstandings that may harm the ML-enabled system. Indeed, previous works have suggested recommendations for collaboration [7, 12, 13]. However, there is still a gap in assessing their relevance for typical technical tasks in ML-enabled systems based on practitioners' perspectives.

Therefore, our study aims to understand how to enhance the collaboration between SWEs and DSCs when developing ML-enabled systems. For this purpose, we defined two research questions: *"RQ1) What is the perception of SWEs and DSCs on which technical tasks they judge as most relevant to collaborate?"* and *"RQ2) What is the perception of SWEs and DSCs on the relevance of suggested recommendations to enhance collaboration on technical tasks?"*. The first question allowed us to understand which tasks are considered most relevant to collaborate. The second enabled us to analyze the suggested recommendations' relevance for the investigated tasks.

To answer our questions, we designed two Focus Group (FG) sessions with experienced SWEs and DSCs working on ML-enabled systems. Following the guidelines proposed by Kontio *et al.* [10], we promoted enriched discussions to examine perceptions of SWEs and DSCs on which tasks they consider most relevant to collaborate during ML-enabled systems development. Furthermore, participants were asked to evaluate a curated list of recommendations suggested by the literature to improve collaboration.

Regarding the contributions and novelty, our study advances the understanding of how collaboration between SWEs and DSCs is essential for building ML-enabled systems. Specifically, the collaboration can considerably benefit tasks like defining data access and integrating the ML model with other system components. Our findings also highlight the importance of clear role boundaries and ensuring that all project stakeholders possess a certain level of ML literacy to enhance collaboration. With respect to our work's relevance, the implications of our research can be helpful for organizations seeking to leverage the collaboration and performance of their teams responsible for developing ML-enabled systems.

This paper is structured as follows. Section 2 concerns the background and related work. Section 3 outlines our method. Section 4 details the results. Section 5 discusses the answers to our research questions. Section 6 concludes the paper and suggests future work.

## 2 BACKGROUND AND RELATED WORK

The literature confirms how developing ML components challenges current collaboration practices. For example, Wan *et al.* [17] conducted a mixed study comparing ML and non-ML systems to grasp their most important differences according to practitioners. The authors noted that ML systems rely heavily on data experimentation. This dependency, in turn, introduces uncertainties when establishing system requirements and estimating task effort. Given this scenario, the authors highlighted that communication between team members is vital. Moreover, Amershi *et al.* [3] presented a case study with Microsoft software development teams to gather best practices for ML systems. Results showed that "collaboration and working culture" was consistently cited as an important challenge by the participants, regardless of their experience with AI. The authors also clarified that SWEs with traditional systems development knowledge must be able to work alongside ML specialists.

Previous studies also focused on collaborative work from the perspective of DSCs. Kim *et al.* [9] surveyed 793 Data Science (DS) employees and enthusiasts at Microsoft to uncover their work activities and obstacles. While respondents pointed out challenges related to data characteristics, a separate category of challenges was dedicated to team interaction. The authors emphasized the importance of clarifying the goals of the project together with the whole team. In another study, Begel and Zimmermann [5] asked 1500 Microsoft SWEs what questions they would most like DSCs to answer, which were subsequently prioritized by 2500 Microsoft engineers in another survey. While grouping the first survey questions into categories, the authors saw the need to create a category dedicated to collaboration, as participants were interested in practices that could improve the interaction within and between teams. Collaboration also appeared in the results of the second survey. The question "How can we improve collaboration and sharing between teams?" was highly ranked by multiple respondents.

We also found papers investigating collaboration and communication inside multidisciplinary software teams. Zhang *et al.* [18] designed a survey to investigate how professionals working on DS teams collaborate. This survey was answered by 183 IBM employees, including SWEs and DSCs. The findings indicated that DSCs strongly participate in all stages of DS projects, which suggests they may be responsible for guiding the team's activities. In our work, we intend to extend this research scope by examining collaboration between SWEs and DSCs for ML-enabled systems development.

In addition, Lewis *et al.* [11] examined the consequences of ML mismatches between DSCs, SWEs, and operations staff building ML-enabled systems. The authors defined ML mismatches as problems caused by inaccurate assumptions these actors had about the system that could have been prevented through knowledge sharing. Their findings confirmed the existence of mismatches in the collaboration between SWEs and DSCs during product-model integration.

Recently, Mailach and Siegmund [12] discussed how to handle socio-technical challenges for bringing ML-enabled software into production. The authors identified 17 antipatterns, most of them caused by organizational characteristics. The findings reported that tension, communication issues, and difficulties during model integration and deployment characterized the interaction between SWEs and DSCs. Also, Nahar *et al.* [13] interviewed 45 participants working with ML projects to identify challenges and recommendations for the interaction between SWEs and DSCs. They identified three activities that required collaboration: identifying and decomposing requirements, negotiating training data quality and quantity, and integrating DS and SE work. During these tasks, participants reported challenges such as DSCs working isolated from SWEs, insufficient system documentation, and problems with responsibility sharing. Lastly, Busquim *et al.* [7] depicted a case study with a team developing an ML-enabled system to understand collaboration dynamics. After conducting interviews with the team's SWEs and DSCs, the authors uncovered several obstacles that harmed their collaboration. These drawbacks include differences in technical expertise, imprecise definitions of the actors' duties, and the absence of updated documentation.

In this section, we explored papers that addressed collaboration aspects faced by multidisciplinary teams working on ML-enabled systems. While the existing literature proposes recommendations for enhancing collaboration in this context [7, 12, 13], there is a notable research gap in evaluating these recommendations from practitioners' perspectives regarding typical technical tasks. Addressing this opportunity is both relevant and novel for SE research, as it contributes to understanding collaboration for ML-enabled systems development through the lens of SWEs and DSCs.

## 3 METHOD

This section covers our goal and research questions. Then, we detail our data collection and data analysis procedures.

### 3.1 Goal and Research Questions

We used the GQM template [4] to define our study's goal: **Analyze** technical tasks and recommendations suggested by the literature **for the purpose of** enhancing collaboration between SWEs and DSCs **with respect to** the perception of relevance **from the point of view of** experienced SWEs and DSCs **in the context of** developing industrial ML-enabled systems. Based on this research goal, we established two major research questions, as clarified below.

**RQ1: What is the perception of SWEs and DSCs on which technical tasks they judge as most relevant to collaborate?** Rationale: This question focuses on how SWEs and DSCs evaluate the relevance of their collaboration while developing ML-enabled systems. To address this, we reviewed the current literature and identified crucial technical tasks in ML-enabled systems development for participants to discuss.

**RQ2: What is the perception of SWEs and DSCs on the relevance of suggested recommendations to enhance collaboration on technical tasks?** Rationale: This question explores how to enhance collaboration between SWEs and DSCs working on ML-enabled systems. To this end, we curated recommendations suggested by the literature and asked participants to evaluate their relevance for collaboration for each technical task under analysis.

### 3.2 Data Collection

We designed our FG sessions following the guidelines proposed by Kontio *et al.* [10]. In summary, FG is a qualitative research method that involves collecting data through group discussions on a given topic. FG provides truthful and insightful information as it depicts the perception of participants who possess knowledge of the discussed topic [10]. Also, FG has been used in other SE studies to gather developers' perspectives [2]. As we were looking for insights from practitioners, we considered FG a suitable choice.

*3.2.1 Focus Group Design.* Our first step towards designing the FG was to define how it would be conducted. We had to specify what topics would be discussed and how to foster this discussion. We organized our FG sessions in two stages: one for debating tasks related to collaboration between SWEs and DSCs, and one for examining recommendations for improving this collaboration.

Regarding the **first stage**, we had to decide which technical tasks related to developing ML-enabled systems would be most suitable for discussion. To this end, we extracted tasks depicted as important collaboration points between SWEs and DSCs from different works [7, 11–13, 16]. These works were chosen because they explicitly report collaborative aspects of developing ML-enabled systems. Following this analysis, a total of seven tasks have been identified:

<u>Data Access Definition</u>: Acquiring the necessary data to develop and evaluate an ML model may require collaboration between multiple parties [13].

<u>Data Selection</u>: Selecting the data used to build the model and describing what each data component represents. SWEs reported that knowing the meaning of the data fields used by their respective ML models could have prevented errors in the system [7].

<u>ML Model Evaluation</u>: The evaluation of an ML model involves more than just measuring its performance using metrics. It also takes into account aspects like the model's interpretability. Kim *et al.* [9] have shown that team members may lack knowledge about the model, which is why we agreed to analyze the importance of collaboration during this task.

<u>ML Artifact Storage</u>: The systematic storage and management of all ML artifacts, including scripts and the model itself. Since this topic may involve both SWEs and DSCs [7], we aim to determine whether improved communication between these groups can enhance the storage of ML artifacts.

<u>ML Model Availability</u>: Make the ML model accessible to other system components so that they can use it. Busquim *et al.* [7] exemplified a dissatisfaction with DSCs being solely responsible for this task. They lacked the necessary skills to perform it and had to seek assistance from SWEs.

<u>ML Model Integration</u>: Integrating the ML component into the larger software system. Previous works have identified several challenges related to collaboration during this process, such as data being in the wrong format during integration tests [11], communication problems [12], and a lack of proper documentation for component integration [13].

<u>ML Model Deployment</u>: Making the ML model available in a production environment. DSCs may require SWEs' assistance with model infrastructure issues [13].

On the other hand, the **second stage** of our FG aimed to promote the debate around recommendations suggested by the literature to improve collaboration between SWEs and DSCs. To determine which recommendations would be discussed, we analyzed the works of Nahar *et al.* [13], Mailach and Siegmund [12], and Busquim *et al.* [7]. We selected these three studies based on their explicit proposals for improving collaboration in the context of ML-enabled systems. After reviewing the papers, we extracted the most relevant recommendations and compiled them into a final list of six suggestions. Each one was assigned a reference ID, which we will use to refer to them in our analysis.

***R1) Involve DSCs and business owners when eliciting and analyzing requirements***: During the requirements definition stage, it may be helpful to have DSCs, business owners, and SWEs interact to clarify requirements and discard unrealistic ones [12, 13].

***R2) Provide ML literacy for all project stakeholders***: Aims to establish a shared understanding of basic ML concepts/terminologies across all team members [12, 13].

***R3) Develop documentation for product requirements, system architecture, and APIs at collaboration points***: Encourages producing documentation that covers the whole ML-enabled system development process [7, 12, 13].

***R4) Define clear responsibilities and internal processes with clear boundaries for SWEs and DSCs***: Clarifies the roles and responsibilities of both SWEs and DSCs within the team, so that they can understand their tasks and execute them successfully [7, 12, 13].

***R5) Support interdisciplinarity between SWEs and DSCs***: Fosters knowledge sharing between actors and provides opportunities for both groups to learn about each other's work [7, 12, 13].

***R6) Organize regular meetings for showcasing team activities***: Aims to improve collaboration within the team by arranging meetings that involve all members. During these meetings, team members can showcase their work, share their findings, coordinate tasks, and communicate important messages to everyone [7, 12, 13].

*3.2.2 Participant Recruitment.* Once we completed the design of our FG, we began the process of recruiting participants. We utilized convenience sampling to reach out to fourteen professionals currently working on ML-enabled system projects. Out of the fourteen emails sent, we received seven positive responses. Each participant who confirmed their attendance was sent a consent and characterization form to fill out. The consent form clearly stated the objective of the FG, including its estimated duration time and risks. We assured participants that all data collected would be anonymized, kept confidential, and used solely for research purposes.

We asked participants about their educational qualification, work experience, and the number of projects they worked on with ML components. We also inquired if they identified themselves as a DSC or a SWE and about their proficiency in both areas. To evaluate their competency, we used the NIH Proficiency Scale[1], which is a widely-known tool to assess proficiency. This scale has six proficiency levels: Not Applicable, Fundamental Awareness, Novice, Intermediate, Advanced, and Expert. We included each level and its description in the form to ensure participants' understanding.

*3.2.3 Operationalization of the FG sessions.* The FG sessions were conducted online using Google Meet. Based on the number of participants, we organized two separate sessions to diversify the shared experiences. To present the topics for discussion and enable exchange between participants, we created interactive boards inside Miro[2]. Figure 1 illustrates the upper part of the board created for the first stage of the FG. The board contains all technical tasks described in Section 3.2.1. The participants were asked to rate the relevance of collaboration between SWEs and DSCs for each task. The rating scale ranged from agreeing to disagreeing, and participants could also choose not to express an opinion. To cast their votes, participants had to drag their post-it notes to the desired answer for each task displayed on the board.

---

[1] https://hr.nih.gov/working-nih/competencies/competencies-proficiency-scale
[2] https://miro.com

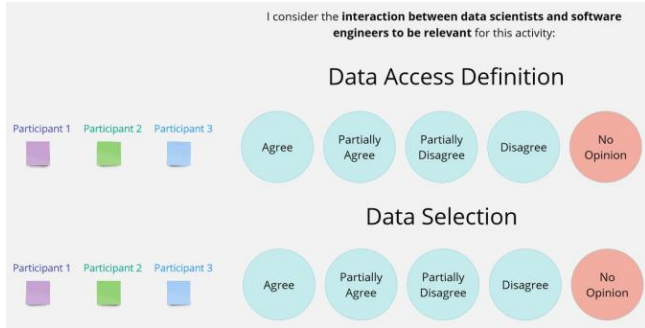

**Figure 1: Miro Board for the First Stage of the FG**

Two authors acted as moderators in both FG sessions. In the **first stage** of the FG, they described the research goal and read out loud each of the tasks that would be examined. Participants were allowed to ask questions about the tasks if they had doubts, which were answered through examples. After that, they had one minute to cast votes for all tasks. When voting was finished, participants had seven minutes to discuss the thought process behind their answers. During the discussion, they could also change their votes in case they felt persuaded by other participants. Relevant comments made by the participants were registered on the board by one of the moderators. The moderators also asked clarifying questions to the participants when there was a need for better understanding.

After the end of the discussion, the moderators explained the **second stage** of the FG. In this stage, participants had to examine recommendations for collaboration between SWEs and DSCs. We created six Miro boards for this stage, one for each recommendation. Figure 2 depicts the upper part of the board for the first recommendation. We asked participants to assess the recommendations' relevance for collaboration between SWEs and DSCs for each task. We used the same voting options as in the previous stage.

One of the moderators explained each recommendation and answered questions in case participants had any. Then, participants had to cast their votes for each task. After the last participant's vote, the group had seven minutes to justify their decisions and discuss the relevance of that recommendation for collaboration. Once again, the moderators intervened whenever a comment needed more context to be properly understood. When the discussion was over, the moderators advanced to the next recommendation until all recommendations had been examined.

### 3.3 Data Analysis

Both FG sessions lasted one hour and fifteen minutes. All of them were recorded and transcribed with Google Cloud's Speech-to-Text API[3]. During data analysis, we first reviewed and adjusted the transcripts while listening to the recordings. We did this since we noticed some words were not correctly transcribed, and some sentences lacked punctuation. We then removed references to participants' and companies' names to ensure anonymity. The artifacts underpinning this study (including forms, boards, and revised transcriptions) are openly available via our supporting repository [6].

---
[3] https://cloud.google.com/speech-to-text

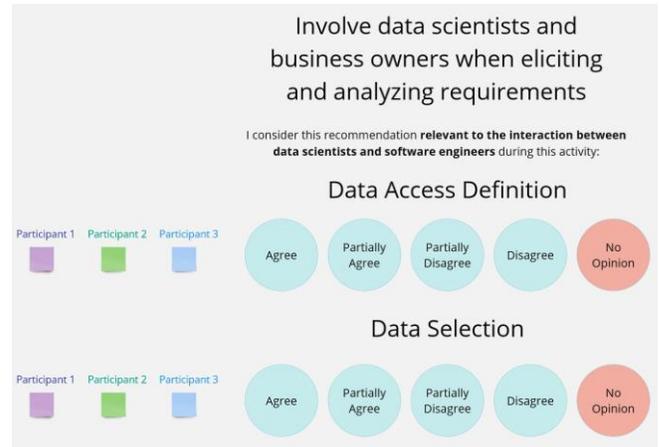

**Figure 2: Miro Board for the Second Stage of the FG**

We examined the content of the interviews using open coding [15]. For each FG, we looked for statements depicting previous experiences with collaboration shared by the participants, including their perceptions regarding the relevance of the recommendations. We also compared the generated codes with the notes registered by the moderators during the sessions to ensure all important topics had been covered. Assigning codes to these statements allowed us to group similar perspectives. Moreover, it enabled us to contrast perceptions from participants of different FG sessions. In the end, we acquired 29 codes, which we used to structure Section 4.

## 4 RESULTS

We assigned identification numbers to the participants to preserve their anonymity. Table 1 displays their demographic data. Participants DSC1, DSC2, DSC3, DSC5, SWE1, and SWE2 identified as male, while DSC4 identified as female. Since only two participants identified as SWEs, we allocated them to different FG sessions. The first session comprised DSC1, DSC2, and SWE1, while the second comprised DSC3, DSC4, DSC5, and SWE2.

**Table 1: Participants' Characterization**

| Participant ID | Role | Educational Qualification | Years of Experience | Number of Projects with ML |
|---|---|---|---|---|
| DSC1 | DSC | Doctorate | 10 | 10 |
| DSC2 | DSC | Doctorate | 5 | 6 |
| DSC3 | DSC | Doctorate | 21 | 11 |
| DSC4 | DSC | Master's degree | 6 | 2 |
| DSC5 | DSC | Doctorate | 7 | 5 |
| SWE1 | SWE | Master's degree | 20 | 2 |
| SWE2 | SWE | Master's degree | 15 | 1 |

Table 2 displays the participants' responses regarding their DS and SE proficiency levels. All participants reported having at least an intermediate level of proficiency in their respective fields, and some even reported proficiency in both areas. All participants have at least five years of work experience and have worked on at least one ML project, making them suitable for our study.

**Table 2: Participants' Proficiency Data**

| Participant ID | DS Proficiency | SE Proficiency |
|---|---|---|
| DSC1 | Advanced | Advanced |
| DSC2 | Advanced | Novice |
| DSC3 | Expert | Advanced |
| DSC4 | Intermediate | Not Applicable |
| DSC5 | Intermediate | Novice |
| SWE1 | Not Applicable | Expert |
| SWE2 | Not Applicable | Expert |

In the following subsections, we describe how participants evaluated the relevance of collaboration for each task and the assessments conducted for each recommendation. The statements depicted in each subsection were selected based on the codes we extracted during the qualitative analysis. Most codes indicate how collaboration may not be relevant for all tasks and the importance of the proposed recommendations. These topics were frequently mentioned throughout both FG sessions.

## 4.1 Relevance of Collaboration for each Task

Table 3 shows how the participants assessed the relevance of collaboration between SWEs and DSCs for each investigated task.

**Table 3: Relevance of Collaboration for each Task**

| Task | Participants who Agreed | Participants who Partially Agreed | Participants who Partially Disagreed | Participants who Disagreed |
|---|---|---|---|---|
| Data Access Definition | DSC1,DSC2, DSC3,DSC4, DSC5 | SWE1,SWE2 | - | - |
| Data Selection | - | - | DSC1,DSC2, DSC3,DSC4, DSC5 | SWE1,SWE2 |
| ML Model Evaluation | - | - | DSC3,DSC4, SWE2 | DSC1,DSC2, DSC5,SWE1 |
| ML Artifact Storage | DSC1,DSC4, DSC5,SWE1 | DSC2,DSC3 | SWE2 | - |
| ML Model Availability | DSC1,DSC2, SWE1 | DSC3,DSC4, DSC5,SWE2 | - | - |
| ML Model Integration | DSC1,DSC2, DSC3,DSC4, DSC5,SWE1, SWE2 | - | - | - |
| ML Model Deployment | DSC1,DSC2, DSC3,DSC4, DSC5,SWE1 | SWE2 | - | - |

Participants mainly agreed that collaboration during the definition of data access is important. DSC1 explained: "*SWEs usually have a greater knowledge of the company's APIs for capturing data, so they should be able to help the DSCs navigate the available infrastructure better. However, this also depends on the role of the DSC. For example, there are teams where the DSC is almost a database administrator as well. In this case, a SWE may not be needed.*" DSC2 stated: "*Considering that data may not be in the hands of the DSC, a SWE can help understand the best way to access it, when to access it, etc.*" During the discussion, SWE2 raised another point concerning this task: "*In the project I am currently working on, we use a collection of documents to train the ML model. These documents were sent to us by customer representatives who are not SWEs, so there was no interaction between these roles. This is why I only partially agreed.*"

Overall, participants tended to disagree with the statement when discussing data selection and model evaluation. SWE1 explained his perception: "*In my previous work experiences, the DSCs were responsible for evaluating the model and selecting data. In my current project, I have no idea how our ML model was evaluated or how data was selected.[...] Still, since we have a good relationship with the DSCs, we are always willing to help them if they need it.*" DSC2 agreed with SWE1's opinion: "*I noticed SWEs are usually not very interested in the research process of a DS application. I have made presentations showcasing the algorithms examined to build a model, or the metrics used for its evaluation, and their attention goes away very quickly.*"

DSC3 explained that selecting data is more suited to the DSC's role: "*The role of a DSC consists of analyzing the data and performing feature engineering to train the model. For these activities, I would not involve a SWE. On the other hand, during data selection, it is possible to deal with incomplete data that you might have to discard or adjust. You may even discover the existence of more data that you still need to acquire. This can lead to a change in how data is being accessed, which may provoke an interaction with a SWE.*" In this sense, DSC1 exemplified how an interaction with a SWE would not necessarily be useful during model evaluation: "*[...] The DSCs know what metrics are relevant. They will know, for example, if the model is overfitted. If you ask SWEs to deploy an overfitted model, they will probably do it, but only a DSC will realize that the model has a problem and is not ready for production. Hence, I do not know how a SWE could help in this process.*"

The majority of the participants reached a consensus that collaboration would be important for storing ML model artifacts. DSC5 justified his vote: "*It is important that both actors define where and how this storage will occur. This interaction with the SWEs allows the DSCs to understand what infrastructure is currently used for storage, as they are usually not involved in this process.*" SWE1 highlighted this interaction is important due to the complexity of the ML component: "*Planning model storage is a task which SWEs should participate. The model can have multiple artifacts, some of which may be enormous, so they must be stored accordingly.*" Finally, DSC3 pointed out another advantage: "*Throughout development, it is possible that team members change, which may harm the improvement of existing models. This can be mitigated by having an infrastructure where artifacts can be properly managed with adequate versioning.*"

Furthermore, there was no disagreement regarding making the ML model available and integrating it with the rest of the system. DSC2 summarized: "*These tasks will define the ML model's output for other system components. This should be discussed between the SWEs and the DSCs, who must evaluate this according to the system's goals.*" Model deployment was another activity where collaboration was viewed as important by DSC1: "*The actors in charge of CI/CD operations are usually SWEs. However, this does not mean they know how to execute the model. For this reason, it is vital that a DSC gets together with them to explain the model and how to run it.*"

## 4.2 R1) Involve DSCs and business owners when eliciting and analyzing requirements

Table 4 depicts how the participants evaluated the relevance of R1 for the collaboration between SWEs and DSCs for each task.

**Table 4: Relevance of R1 for each Task**

| Task | Participants who Agreed | Participants who Partially Agreed | Participants who Partially Disagreed | Participants who Disagreed |
|---|---|---|---|---|
| Data Access Definition | DSC1,DSC2,DSC3,DSC4,DSC5,SWE1 | SWE2 | - | - |
| Data Selection | DSC1,DSC2,SWE1 | - | DSC3,DSC4,DSC5 | SWE2 |
| ML Model Evaluation | DSC1,DSC2 | DSC3,DSC4,DSC5,SWE1 | - | SWE2 |
| ML Artifact Storage | - | DSC1,DSC3,DSC5 | SWE2 | DSC2,DSC4,SWE1 |
| ML Model Availability | - | DSC1 | DSC3,DSC4,DSC5,SWE2 | DSC2,SWE1 |
| ML Model Integration | - | DSC1,DSC3,DSC5 | DSC4,SWE2 | DSC2,SWE1 |
| ML Model Deployment | - | DSC1,DSC3,DSC5 | SWE2 | DSC2,DSC4,SWE1 |

Participants agreed that R1 is important during data access definition. DSC2 explained: "*Involving the business owners can help the actors understand what the data represents. Sometimes, you know the name of a given variable and whether it is numeric or categorical, but you may not understand what it represents. [...] I have worked on projects where the data came from another company. In these cases, the business owners had to explain what each data field represented, and this improved the interaction between us and the SWEs.*" SWE1 agreed with DSC2: "*The business owner knows the data very well. A business owner usually knows how to access the data, whether it is from other systems or from a spreadsheet. I think this interaction is important, especially when defining how data will be acquired and selected.*" DSC3, on the other hand, did not consider this recommendation relevant during data selection: "*I do not think the interaction between DSCs and SWEs is important during this task. The presence of other actors, such as business owners, is more important.*"

DSC3 also commented in favor of R1 for ML model evaluation: "*Everyone must comprehend what exactly will be evaluated. Performance can be evaluated not only in terms of accuracy and other metrics but also in terms of computational performance. There is no point in having a super complex model if it will require a machine with tons of computational power that will not be available. These definitions can sometimes be made together with business owners and SWEs.*" DSC1 had a similar opinion regarding this task: "*Business owners understand a lot about the data and can help with model evaluation. For instance, sometimes you may think the model has to avoid false positives, but they might say, 'No, my problem is with false negatives,' so then you will have to choose another metric.*" DSC1 also gave an example of when this recommendation would be useful: "*Having the business owners close to the SWEs and DSCs during requirements analysis can help enhance their collaboration. Imagine a scenario where the business owner asks for a given accuracy and latency. The DSC knows how to achieve that accuracy, but may not know if that latency is possible given the available machines. This is something a SWE can help with.*"

The participants had different opinions regarding ML model integration and deployment. They either partially agreed, partially disagreed, or disagreed entirely. DSC4 explained her votes for these tasks: "*I do not think this recommendation would be that interesting for these tasks. [...] I consider model integration more of an implementation task, just like with deployment, so perhaps the business owners might not be needed.*" DSC3 had a different opinion: "*In my view, any prioritization or decision-making often involves business owners. In some cases, the integration task may involve different teams, and the business owner will be able to facilitate this coordination.*"

## 4.3 R2) Provide ML literacy for all project stakeholders

Table 5 shows how our participants evaluated the relevance of R2 for each task.

**Table 5: Relevance of R2 for each Task**

| Task | Participants who Agreed | Participants who Partially Agreed | Participants who Partially Disagreed | Participants who Disagreed |
|---|---|---|---|---|
| Data Access Definition | DSC5 | DSC3,DSC4,SWE2 | - | DSC1,DSC2,SWE1 |
| Data Selection | - | - | DSC1,DSC3,DSC4,DSC5,SWE2 | DSC2,SWE1 |
| ML Model Evaluation | - | DSC3,DSC4,DSC5,SWE2 | - | DSC1,DSC2,SWE1 |
| ML Artifact Storage | DSC2 | DSC1,DSC3 | DSC5,SWE1,SWE2 | DSC4 |
| ML Model Availability | DSC2 | DSC1,DSC3,DSC5,SWE2 | DSC4,SWE1 | - |
| ML Model Integration | DSC4,DSC5,SWE2 | DSC1,DSC2,DSC3 | SWE1 | - |
| ML Model Deployment | - | DSC1,DSC3,DSC4,DSC5,SWE2 | DSC2,SWE1 | - |

DSC1 did not see this recommendation relevant while defining data access: "*During data access definition, I do not think knowing the difference between classification and regression would be helpful for acquiring the data. [...] However, if SWEs happen to know a bit more about DS, they may be able to help with data selection. For example, they may discover noise in the data capable of hindering model training, or notice that a data column has many null values.*" SWE1 explained how his work experience influenced his view on the effect of this recommendation during data access definition: *I have been working in an academic environment for a long time. Even though I am not in the AI field, I constantly see lectures and learn about this topic, so I did not require this literacy. Having said that, I do not think this theoretical knowledge is important for data access definition. I think practical instructions, such as how to access a spreadsheet or another system, are more efficient.*"

DSC3 assessed R2 positively: "*I consider literacy relevant, as knowing at least the basics is important for communication. The only exception I can see is during data selection because I think the participation of the SWE is reduced. I partially agreed on the other tasks because, in the worst case, everyone has to know that a model will be executed, that an output of a certain type will be generated, etc.*" SWE2 agreed with his opinion: "*This recommendation is important to improve communication between the SWE and the DSC. It is vital to establish a common terminology so that communication flows more easily.*"

DSC3 also made comments regarding the ML artifact storage task: "*As I mentioned [...], it is a good idea to define the best way to store the model together with the SWEs to make it more easily accessible in the future. [...] They must understand what model training is and what is actually being stored.*" DSC1 exemplified this advantage: "*Comprehending how the models are generated allows you to think about how to persist them more intelligently. For example, suppose there is a high chance of data drifts in the project you are working on. In that case, the SWE helping with ML artifact storage would know that the model needs to be trained and updated often. This would determine the development of an effective versioning system.*"

When debating ML model deployment and integration, DSC1 also perceived benefits: "*ML literacy for SWEs greatly helps in these tasks. They will know, for example, what type of approach to adopt when deploying the model. Depending on the type of model the DSC has created, the SWE will have an idea of the computational power required to run it. [...] The SWE will be able to notice this even if the DSC forgets to warn the team.*" DSC2 emphasized the benefits of this recommendation during model integration: "*ML literacy can help with the definition of the model's output and how it will be consumed. It makes communication between the actors easier when specifying the best way to interact with the model.*"

### 4.4 R3) Develop documentation for product requirements, system architecture, and APIs at collaboration points

Table 6 presents how our participants evaluated the relevance of R3 for each task.

**Table 6: Relevance of R3 for each Task**

| Task | Participants who Agreed | Participants who Partially Agreed | Participants who Partially Disagreed | Participants who Disagreed |
|---|---|---|---|---|
| Data Access Definition | DSC3,DSC4, DSC5,SWE2 | - | DSC1,DSC2, SWE1 | - |
| Data Selection | - | DSC1,DSC2, SWE1 | DSC3,DSC4, DSC5 | SWE2 |
| ML Model Evaluation | DSC4 | DSC3,DSC5, SWE2 | - | DSC1,DSC2, SWE1 |
| ML Artifact Storage | DSC4 | DSC1,DSC3, DSC5,SWE2 | - | DSC2,SWE1 |
| ML Model Availability | DSC2,DSC4, DSC5,SWE1 | DSC1,DSC3, SWE2 | - | - |
| ML Model Integration | DSC2,DSC3, DSC4,DSC5, SWE1,SWE2 | DSC1 | - | - |
| ML Model Deployment | DSC2,DSC3, DSC5,SWE1 | DSC1,DSC4, SWE2 | - | - |

Some participants did not consider this recommendation relevant for certain tasks. One of them was DSC3, who explained: "*I do not think this recommendation is relevant for data selection because I do not see the need for much interaction between SWEs and DSCs during this task. For ML model evaluation, I think this interaction exists, but it is not strong, so I partially agreed with the statement.*" DSC4, on the other hand, shared a different opinion regarding model evaluation: "*The model will be evaluated based on the documented requirements, so I think this recommendation has a lot of influence on this process.*"

All participants either agreed or partially agreed with the statement for model availability, model integration, and model deployment. DSC1 illustrated his point of view, also shared by other participants: "*Having well-produced documentation greatly impacts development. For example, if I have a requirement for extremely low latency, it may affect how the model will be developed and made available. The fact that this is documented explicitly assists in the collaboration between the SWE and the DSC. [...] If the DSC comes up with an extremely slow approach, it will be clear to everyone that the model is not ready for production.*" DSC5, a DSC, emphasized the importance of this recommendation during the interaction with SWEs: "*Especially when defining APIs, this documentation is important for validating what will be done with the SWEs.*"

### 4.5 R4) Define clear responsibilities and internal processes with clear boundaries for SWEs and DSCs

Table 7 illustrates how participants evaluated the relevance of R4 for each task.

**Table 7: Relevance of R4 for each Task**

| Task | Participants who Agreed | Participants who Partially Agreed | Participants who Partially Disagreed | Participants who Disagreed |
|---|---|---|---|---|
| Data Access Definition | DSC1,DSC2, DSC3,DSC4, DSC5,SWE1, SWE2 | - | - | - |
| Data Selection | DSC1,DSC2, DSC3,DSC4, DSC5,SWE2 | SWE1 | - | - |
| ML Model Evaluation | DSC1,DSC2, DSC3,DSC4, DSC5,SWE2 | - | SWE1 | - |
| ML Artifact Storage | DSC1,DSC2, DSC3,DSC4, DSC5,SWE1, SWE2 | - | - | - |
| ML Model Availability | DSC1,DSC2, DSC3,DSC4, DSC5,SWE1, SWE2 | - | - | - |
| ML Model Integration | DSC1,DSC2, DSC3,DSC4, DSC5,SWE1, SWE2 | - | - | - |
| ML Model Deployment | DSC1,DSC2, DSC3,DSC4, DSC5,SWE1, SWE2 | - | - | - |

We can see that almost all participants agreed with the importance of this recommendation for all tasks. DSC3 explained a reason for this: "*This recommendation is important regardless of how much interaction occurs during each task. Even if there is an activity where there is not supposed to be any collaboration between a SWE and a DSC, this must be clear for everyone to avoid someone doing something that is not their responsibility. This recommendation will help define what interactions will happen during the project.*" DSC5 agreed with DSC3: "*Even for tasks with reduced interaction, following this recommendation guarantees everyone knows their role and what they must do.*" During the discussion, DSC1 emphasized the importance of defining DSCs' responsibilities: "*The DSC's role can often get confused with other roles. They are sometimes also considered database administrators. In addition, they may be confused with SWEs and* *expected to handle model deployment solely. When you clearly define each role's responsibility, communication becomes easier.*"

However, SWE1 partially disagreed with the relevance of this recommendation during ML model evaluation: "*In the teams I have worked in, explicitly defining responsibilities and boundaries was never necessary during this task. Both DSCs and SWEs already knew what was expected from them without a previous formal explanation. I believe this task should be mostly carried out by DSCs, as there is no need for SWEs.*"Lastly, DSC2 mentioned another advantage promoted by this recommendation: "*Clarifying boundaries helps a lot in communication, especially when evaluating the work we need to do. From the point of view of system architecture, these definitions help us visualize who will be responsible for each system component.*"

### 4.6 R5) Support interdisciplinarity between SWEs and DSCs

Table 8 displays the relevance assigned to R5 by the participants for each task.

**Table 8: Relevance of R5 for each Task**

| Task | Participants who Agreed | Participants who Partially Agreed | Participants who Partially Disagreed | Participants who Disagreed |
|---|---|---|---|---|
| Data Access Definition | DSC2,SWE1 | DSC1,DSC5 | DSC3,DSC4,SWE2 | - |
| Data Selection | SWE1 | DSC1,DSC2 | DSC3,DSC4,DSC5,SWE2 | - |
| ML Model Evaluation | SWE1 | DSC1,DSC2 | DSC3,DSC4,DSC5,SWE2 | - |
| ML Artifact Storage | DSC1,DSC2,SWE1 | - | DSC3,DSC4,DSC5,SWE2 | - |
| ML Model Availability | DSC1,DSC2,SWE1 | - | DSC4,SWE2 | DSC3,DSC5 |
| ML Model Integration | DSC1,DSC2,SWE1 | DSC5,SWE2 | DSC3,DSC4 | - |
| Deploy the ML model | DSC2,SWE1 | DSC1,DSC3 | DSC4,DSC5,SWE2 | - |

DSC3, DSC4, SWE2, and DSC5 partially disagreed with this recommendation's relevance for most tasks. They shared a similar view, as explained by DSC4: "*Knowing about other fields is always beneficial, but I do not consider this essential for any activity.*" DSC5 mostly agreed with DSC4, except for some tasks: "*During data access definition, having both actors working closely may speed up the process. This recommendation can also help during model integration, as this task requires a lot of collaboration.*"

DSC1 agreed and perceived benefits from this recommendation: "*For data access definition, data selection, model evaluation, and model deployment, interdisciplinarity is interesting but not vital. A lack of knowledge exchange between the actors during these tasks would not threaten the project. [...] I consider this recommendation important for the other tasks. For artifact storage [...], both actors need to know how the model was developed and what should be versioned. The same goes for model availability and integration: how to consume the model and its inputs and outputs must be clear to everyone.*"

DSC2 described how acquiring SE skills could be interesting for a DSC: "*DSCs are usually more interested in research and generating insights through data analysis, so they may not know how a solution can actually be operationalized and sustained over time. Before deploying to production, it is important to understand what infrastructure is available, how the model will be updated, and how data will be continuously obtained.*" DSC2 also highlighted this knowledge exchange could also be beneficial for SWEs: "*After the system is deployed and operational, the actor responsible for maintaining it is usually a SWE, not a DSC. This recommendation motivates the SWE to know more about model characteristics.*"

### 4.7 R6) Organize regular meetings for showcasing team activities

Table 9 illustrates how our participants evaluated the relevance of R6 for each task.

**Table 9: Relevance of R6 for each Task**

| Task | Participants who Agreed | Participants who Partially Agreed | Participants who Partially Disagreed | Participants who Disagreed |
|---|---|---|---|---|
| Data Access Definition | DSC2,DSC3,DSC4,DSC5,SWE1 | DSC1,SWE2 | - | - |
| Data Selection | SWE1 | DSC1,DSC2,DSC3,DSC4,DSC5 | SWE2 | - |
| ML Model Evaluation | DSC3,DSC4,DSC5 | DSC1,DSC2,SWE1,SWE2 | - | - |
| ML Artifact Storage | DSC1,DSC2 | DSC3,DSC4,DSC5,SWE1 | SWE2 | - |
| ML Model Availability | DSC1,DSC2 | DSC3,DSC4,DSC5,SWE1 | SWE2 | - |
| ML Model Integration | DSC1,DSC2,DSC3,DSC4,DSC5,SWE2 | SWE1 | - | - |
| ML Model Deployment | DSC1,DSC2,DSC3,DSC4,DSC5 | DSC1,DSC2,SWE1,SWE2 | - | - |

Almost all participants either agreed or partially agreed with the relevance of this recommendation. DSC3 explained his view: "*The interaction between SWEs and DSCs will be greatly facilitated if you know exactly what each person is doing and what is happening in the project. This recommendation is valid for monitoring the team and exchanging knowledge, as you can even schedule technical meetings if needed. [...] These regular meetings to find out how things are going and what is being done already help a lot.*" SWE1 stressed this recommendation makes more sense for some tasks than others: "*When defining data access and selecting data, I think the meetings would help. For the other tasks, I consider regular meetings useful, but not critical. After the team has agreed on all definitions, an occasional conversation between members should be enough.*"

DSC1 had a different opinion: "*I love daily meetings because they bring team members from different areas together to witness the whole project's evolution. For tasks that require a greater synergy between SWEs and DSCs, I agree that there should be regular meetings to keep everyone up to date. However, for tasks where I do not see such a strong synergy, I believe these meetings are nice to have [...], but they are not crucial. In the case of data selection [...], a meeting to discuss this task would be good for the SWEs to be more informed about what is happening, but it is not essential for their work. The same is true for data access definition. Showcasing how data is obtained might be interesting for DSCs, but they do not have to know where the data comes from as long as they have access to it.*"

## 5 DISCUSSION

This section portrays a discussion of the results presented in this paper, focusing on how they relate to the two research questions that guided our study.

*RQ1: What is the perception of SWEs and DSCs on which technical tasks they judge as most relevant to collaborate?* Collaboration plays an important role during the definition of data access. SWEs can help DSCs acquire the data needed to develop and validate the ML model, as they usually know how to access the available APIs and databases. However, there may also be cases where the DSCs obtain the data independently. This happens when data is received manually by another actor, such as a customer representative, or when DSCs are also database administrators. In these scenarios, an interaction with a SWE may not be required.

Data selection is usually a DSC's responsibility [9], which makes collaboration with SWEs less frequent during this task. Moreover, participating in this process may sometimes seem uninteresting for the SWEs, which contributes to less interaction. Yet, during this task, a DSC may come across incomplete data for training the model. If this leads to any modifications in data access definitions, then a SWE may be involved. Collaboration between SWEs and DSCs may also not be needed for ML model evaluation, as DSCs should be able to perform this task independently. Once again, SWEs might lack interest in understanding the metrics used for evaluation. Since they usually do not possess this knowledge, it might not be valuable to interact with them during this task.

Collaboration between the actors can benefit the team when specifying ML artifact storage. Creating and maintaining an adequate storage infrastructure for the model is a challenge, given the large size of some ML artifacts and the need to manage their different versions [7]. This difficulty can be mitigated by guaranteeing that SWEs and DSCs are involved when defining how and where each artifact will be stored. Our findings reinforce that collaboration is fundamental for assessing ML model availability and integration, since these tasks relate to the interaction of the ML model with other system components. They include defining how the model will be consumed and what output it will provide [16]. For this reason, performing these tasks implies several discussions between SWEs and DSCs. These discussions are also relevant during model deployment. Even though SWEs usually handle deployment tasks, DSCs should also be present to explain the ML model and how it should be executed. Having them next to the SWEs can also help resolve latency issues caused by the model.

> **RQ1 Answer:** Collaboration between SWEs and DSCs is fundamental for several technical tasks, as their interaction can greatly help face the challenges involved in developing ML-enabled systems. In particular, tasks that can benefit from this interaction include specifying data access and integrating the ML model. For other tasks, such as data selection and ML model evaluation, however, the team must analyze if promoting collaboration would be more beneficial than assigning a single actor as responsible.

*RQ2: What is the perception of SWEs and DSCs on the relevance of suggested recommendations to enhance collaboration on technical tasks?* Participants raised important features of each recommendation we selected. The fact they were analyzed in the context of specific tasks allowed respondents to give practical examples to support their points of view. Involving DSCs and business owners when eliciting and analyzing requirements can be very useful when defining how data will be accessed. Business owners are usually familiar with the data and can explain what they represent and how to access them correctly. Sharing this knowledge with the team can improve the communication between SWEs and DSCs when discussing this topic. Model evaluation can also benefit from this recommendation, as business owners can use their understanding of the data to define the most suitable performance metrics. For the other tasks, the influence of this recommendation will depend on the project's context. For example, when different teams are involved in ML model integration, having a business owner coordinating them during requirements analysis may be helpful.

Providing ML literacy for all project stakeholders is important for collaboration because it aids in establishing a common terminology inside the team, making communication more efficient. It also helps team members become familiar with the model being developed, as it is vital that they clearly understand the model's goal and the type of output it generates [14]. Participants argued that this recommendation is valuable for collaboration during model deployment, model integration, and ML artifact storage. However, it might not be as relevant for other tasks. For example, ML literacy may be less useful for data access definition than simply providing practical data acquisition instructions, such as what commands should be used to obtain data or what database should be accessed.

Documentation constitutes an important tool for enhancing collaboration [13]. Having all definitions explicitly stated on a document enables team members to understand what is expected from the system. This allows SWEs and DSCs to evaluate each other's work despite their different fields of expertise. Documentation can also be beneficial when the actors have to plan their tasks. For model integration and availability, documenting API contracts and model inputs and outputs can help clearly define what each actor must do, making the development process more efficient.

Clearly defining responsibilities and boundaries between SWEs and DSCs is critical, even for tasks that may not require much collaboration. This recommendation raises awareness about the activities each actor is supposed to perform. For example, it must be clear from the beginning to everyone in the team whether DSCs should be responsible for handling ML model deployment. Doing this will avoid problems later in the project and improve their communication with SWEs during this task if needed. This recommendation can also be useful when specifying the system's architecture, as each actor's responsibilities might be directly related to the system components they will have to develop.

Supporting interdisciplinarity between SWEs and DSCs may not be as vital as other recommendations we assessed. If responsibilities are correctly assigned, a lack of knowledge exchange between the actors during the tasks should not compromise the team's performance. Nevertheless, following this recommendation can still be useful since it enables the team to work closely. For tasks such as model deployment and integration, where collaboration may be required, interdisciplinarity allows actors to learn more about each other's work. Besides improving communication, this can also benefit the team in the long term. For instance, DSCs may use this knowledge to develop ML models in such a way that makes their deployment easier in the future.

Organizing regular meetings provides several advantages for team alignment and collaboration. They allow team members to know what is being done and discuss the project's current state. Moreover, these meetings can foster knowledge exchange and help resolve issues during development. However, to improve collaboration, the team must organize these meetings properly. For example, it may not be interesting to schedule meetings with both SWEs and DSCs to discuss tasks that do not require their cooperation. For this reason, teams must evaluate the importance of each meeting based on the project's current state and the team's characteristics.

> **RQ2 Answer:** The relevance of the recommendations for each technical task depends on project characteristics and the level of interaction between SWEs and DSCs during task execution. Most recommendations are relevant for tasks that require strong collaboration, such as ML model integration. Still, recommendations like defining clear responsibilities and organizing regular meetings can be valuable regardless of the level of collaboration.

## 5.1 Threats to Validity

A potential threat to **construct validity** could be related to our research design not being suited to answer our research questions. However, we carefully established the tasks and recommendations discussed during the FG sessions based on the current literature on collaboration for ML-enabled systems. To improve the credibility and representativeness of our results, we selected papers published in prestigious venues with findings that were acquired based on perceptions from professionals working on industry ML projects.

Threats to **internal validity** include participants not understanding the tasks and recommendations we defined for discussion, as well as applying a different methodology in each FG. To mitigate this threat, we followed a standardized procedure during the two FG sessions. Tasks and recommendations were discussed in the same order, and we timed the discussions to ensure all topics were debated equally. Before each voting phase, we allowed participants to ask questions to better understand the task or recommendation they had to assess. While voting, participants could see each other's votes, which can induce conformity bias. We tried to mitigate this by reinforcing that there were no right or wrong answers, as the votes reflected each participant's previous experiences. We encouraged participants to explain the motivation behind their votes.

A threat to **external validity** concerns our study results not being valuable for other teams working with ML-enabled systems. Hence, we recognize the limited number of participants in our FG may be considered a threat to our findings' validity. To mitigate this threat, we invited experienced professionals who work on different ML-enabled systems to participate in our research. According to Debus [8], it is recommended to have at least two sessions for each relevant variable, or until the information gathered is no longer new. Practitioners should examine our results to visualize how they can be applied to their teams. To strengthen the **reliability** of our results, we provide an online repository [6] with all our artifacts.

## 6 CONCLUSION

We have researched the viewpoints of SWEs and DSCs to understand how to enhance their collaborative processes while developing ML-enabled systems. We clarified that different studies have highlighted issues related to the collaboration between these two actors. Although some works have provided recommendations for enhancing such collaboration, we noticed a research gap with respect to assessing the relevance of these recommendations with practitioners. Hence, we conducted two FG sessions with experienced SWEs and DSCs working on ML-enabled systems to acquire their perceptions regarding the relevance of collaboration for technical tasks and how they assessed recommendations for improving their interaction. To do this, we analyzed the literature on collaboration for ML-enabled systems and selected a group of technical tasks and recommendations for the FG discussions. We transcribed all FG sessions and qualitatively analyzed each participant's comments to answer our research questions.

Our findings contribute to researchers and practitioners interested in how collaboration between these actors unfolds during tasks relevant to ML-enabled systems development. We discovered that having SWEs and DSCs collaborating on technical tasks such as ML model integration and deployment can foster knowledge exchange and prevent errors in the system. Moreover, our results contain in-depth discussions and practical examples of the effects of each assessed recommendation. For instance, defining clear responsibilities and boundaries is relevant for collaboration during several technical tasks, as it improves the team's communication. We believe these analyses can help other teams working on ML-enabled systems identify the most appropriate recommendations for enhancing their performance.

Regarding future work opportunities, it would be worthwhile to further explore the recommendations that were evaluated in our FG sessions. For example, future studies could investigate how to apply these recommendations in practical settings and monitor their effects on the teams' productivity. Another opportunity involves conducting additional studies to reinforce our findings and explore additional stakeholders' perspectives.

## ONLINE RESOURCES

All the artifacts underpinning this study are openly available via our supporting repository [6] under the Creative Commons Attribution license, ensuring transparency and accessibility. This allows other practitioners to validate our results and replicate our research design in future studies.

## ACKNOWLEDGMENTS

We thank the Brazilian Council for Scientific and Technological Development (CNPq process #312275/2023-4) and StoneCo Ltd (research project "Software Engineering for Data Science and Artificial Intelligence") for financial support.